# Tensile behavior of dual-phase titanium alloys under high-intensity proton beam exposure: radiation-induced omega phase transformation in Ti-6Al-4V

Taku Ishida [a,b,*], Eiichi Wakai [a,c], Shunsuke Makimura [a,b], Andrew M. Casella [d], Danny J. Edwards [d], Ramprashad Prabhakaran [d], David J. Senor [d], Kavin Ammigan [e], Sujit Bidhar [e], Patrick G. Hurh [e], Frederique Pellemoine [e], Christopher J. Densham [f], Michael D. Fitton [f], Joe M. Bennett [f], Dohyun Kim [g], Nikolaos Simos [g †], Masayuki Hagiwara [b], Naritoshi Kawamura [a,b], Shin-ichiro Meigo [a,c], Katsuya Yonehara [e]

**On behalf of the RaDIATE COLLABORATION**

[a] *Japan Proton Accelerator Research Complex (J-PARC), Tokai-mura, 319-1106, Japan*
[b] *High Energy Accelerator Research Organization (KEK), Tsukuba, 305-0801, Japan*
[c] *Japan Atomic Energy Agency (JAEA), Tokai-mura, 319-1106, Japan*
[d] *Pacific Northwest National Laboratory (PNNL), Richland, WA 99352, U.S.A.*
[e] *Fermi National Accelerator Laboratory (Fermilab), Batavia, IL 60510-5011, U.S.A.*
[f] *Science and Technology Facilities Council (STFC) Rutherford Appleton Laboratory (RAL), Oxfordshire, OX11 0QX, U.K.*
[g] *Brookhaven National Laboratory (BNL), Upton, NY 11973-5000, U.S.A.*

[*] *Corresponding author.*
  *E-mail address: taku.ishida@kek.jp (T. Ishida).*
[†] *Deceased 15/07/2020*

**Abstract**

A high-intensity proton beam exposure with 181 MeV energy has been conducted at Brookhaven Linac Isotope Producer facility on various material specimens for accelerator targetry applications, including titanium alloys as a beam window material. The radiation damage level of the analyzed capsule was 0.25 dpa at beam center region with an irradiation temperature around 120 °C. Tensile tests showed increased hardness and a large decrease in ductility for the dual $\alpha+\beta$-phase Ti-6Al-4V Grade-5 and Grade-23 extra low interstitial alloys, with the near $\alpha$-phase Ti-3Al-2.5V Grade-9 alloy still exhibiting uniform elongation of a few % after irradiation. Transmission Electron Microscope analyses on Ti-6Al-4V indicated clear evidence of a high-density of defect clusters with size less than 2 nm in each $\alpha$-phase grain. The $\beta$-phase grains did not contain any visible defects such as loops or black dots, while the diffraction patterns clearly indicated $\omega$-phase precipitation in an advanced formation stage. The radiation-induced $\omega$-phase transformation in the $\beta$-phase could lead to greater loss of ductility in Ti-6Al-4V alloys in comparison with Ti-3Al-2.5V alloy with less $\beta$-phase.

**Keywords**

Titanium alloy
Omega phase
Radiation damage
Accelerator target
Beam window

## 1. Introduction

Titanium alloys are widely utilized as structural materials for shipbuilding, chemical and aerospace applications, because of their high specific strength, good fatigue endurance limits, and good corrosion/erosion resistance [1,2]. They also satisfy low-activation requirements for use as nuclear power materials. The dual phase alloy, Ti-6Al-4V (α-phase: Hexagonal Close-Packed, HCP; β-phase: Body-Centered Cubic, BCC), is one of the most used titanium alloys, which exhibits remarkably high strength at room-temperature up to 300 °C and excellent fatigue performance. Several high-intensity proton accelerator facilities utilize this alloy for beam windows, which separate accelerator vacuum and target station vessel atmosphere (such as helium or nitrogen). The J-PARC neutrino facility utilizes it for both its primary beam window and target containment window [3,4], and the Long Baseline Neutrino Facility (LBNF), under design at Fermilab, plans to adopt it as a target containment window [5,6]. At these facilities, passage of the high-intensity pulsed proton beam through the beam window causes localized energy deposition. Consequent instantaneous temperature rise and thermal expansion create thermal shock, which propagates in the beam window as dynamic stress waves, and can cause failure. Relatively low density of Ti-6Al-4V is advantageous to reduce the energy deposition, and thus amplitude of the stress waves. The J-PARC primary window [3] comprises two 0.3 mm-thick partial hemispheres of Ti-6Al-4V cooled by helium flowing between the two skins. It experiences $3.3 \times 10^{14}$ protons-per-pulse (ppp) in a few tens of mm$^2$ cross section, and each 5 μs pulse initiates thermal stress waves of a few hundred MPa amplitude. With a planned upgrade to 1.3 MW beam power (repetition rate of about 1 Hz), the window will experience about 8 million beam pulses per each operational year, which corresponds to $2.4 \times 10^{21}$ protons on target (pot). It will accumulate radiation damage of the titanium alloy of about 2 displacement per atom (dpa) per annum at the beam center [7]. Thus, for lifetime prediction of these beam windows, irradiation data of Ti-6Al-4V in the one to a few dpa range are required, and it is particularly important to understand the response of the radiation-damaged titanium alloy to high-cycle thermal shock loading conditions, with greater than 10 million cycles.

Although data on the physical and mechanical properties of Ti-6Al-4V are relatively complete for the non-irradiated condition [8], material property data for titanium alloys in the irradiated condition are limited in general, and are mostly from significantly less than 1 dpa [9]. This is partly because titanium alloys are not being considered for fusion community applications[1] due to the potential for a high tritium inventory [10]. The data that do exist show that titanium alloys are very sensitive to irradiation, even at low doses, exhibiting significant loss of ductility and fracture toughness. Degradation of these properties is severer for the α+β phase alloy than that for single α-phase alloys. Tensile data with low temperature (40~350 °C) irradiation up to 0.3 dpa are available on Ti-6Al-4V and single α-phase Ti-5Al-2.5Sn, for both neutron [11,12] and high energy proton [13,14] irradiations. After irradiation, Ti-6Al-4V specimens showed loss of uniform elongation with the yield strength nearly the same level as the tensile strength, losing any significant ability to work harden. However, after a similar irradiation of Ti-5Al-2.5Sn, significant ductility remained with more than 10% uniform elongation. This different behavior between α+β and α alloys was consistent for both neutron and high-energy proton irradiation data. Similar mechanical tests were made on an α-like Ti-4Al-2V [15] in comparison with Ti-6Al-4V [16], for neutron irradiation up to about 0.4 dpa. Ti-4Al-2V also retained strain hardening capability and uniform elongation did not drop below 2.5%. At a higher temperature (550 °C), very high irradiation data with neutron up to 37 dpa is available for Ti-6Al-4V and two Al/Sn-based near α alloys (6242S and 5621S) [17], showing reasonable post irradiation ductility of more than 8% for both alloys. Post-irradiation TEM analyses reported a high density of defect clusters [11,12] or radiation-induced very fine precipitation [13] uniformly distributed in the α grains. Meanwhile, it is not yet conclusive what microstructure can cause the tensile behavior difference between α+β and α alloys.

---

[1] For ITER, Ti-6Al-4V is considered as flexible support cartridge for blanket attachment system [50,51], where the operational damage dose is expected to be only 0.1 dpa per year.

Recent major accelerator facilities have been limited in beam power by production target and beam window survivability, where radiation damage to their constituent materials has been identified as the most cross-cutting challenge facing these high-power target facilities. To promote these studies, the RaDIATE collaboration, *Radiation Damage In Accelerator Target Environments* [18], was founded to bring together the high energy physics/the basic energy science accelerator target and nuclear fusion/fission materials communities. The research program of the collaboration consists of determining the effect of high energy proton irradiation, which can be very different to that of low energy neutrons, on the mechanical properties of potential target and beam window materials, and to understand the underlying changes by advanced micro-structural studies. The goal is to enable accurate target and window component lifetime predictions, to design robust components for multi-MW beams, and to develop new materials to extend lifetimes.

## 2. Experimental details

Under the aegis of the RaDIATE collaboration, a high-intensity proton beam exposure with 181 MeV energy was conducted at Brookhaven Linac Isotope Producer (BLIP) facility at Brookhaven National Laboratory (BNL) on various material test specimens for accelerator target applications [19], including titanium alloys [20]. Post-Irradiation Examination (PIE) is being conducted at participating reactor/fusion energy research institutions with hot-cell facilities, with Pacific Northwest National Laboratory (PNNL) conducting the work reported here. The principal objective of the studies is to obtain mechanical properties of Ti-6Al-4V, especially the first High Cycle Fatigue (HCF) test data, as current beam window material at J-PARC. A secondary objective is to understand how elemental and phase variations of titanium alloy grades, including commercially-pure titanium, α phase, dual α+β phase, and metastable β phase alloys, affect the irradiation performance. This is to find better radiation-tolerant grades for future upgrades of the beam intercepting devices.

The primary mission of the BLIP facility is to produce medical isotopes using a 112 MeV proton beam. With the BNL Linac's capability to deliver protons up to 200 MeV, it is therefore possible to operate BLIP at a higher beam energy to irradiate accelerator target materials upstream parasitically with the isotope production targets. The target box arrangement in the BLIP beamline is shown in Fig. 1. The RaDIATE target box with six material capsules and one vacuum degrader capsule was installed just upstream of the isotope target box. The total material thickness in the RaDIATE target box was configured accurately in order to degrade the incoming 181 MeV beam and deliver uniform and precise beam energy and fluence to the downstream isotope target box for optimal isotope production. Each material type was enclosed in their individual capsule made of 316L stainless steel, separated in series by a 2.3 mm-wide gap of flowing cooling water. The capsules are hermetically sealed under either inert gas (helium/argon) by laser welding, or vacuum by electron beam welding.

The irradiation campaign on the RaDIATE target box was executed in three phases during 2017 to 2018 for different configurations with six capsules in each phase. In total over 200 specimens in 9 capsules were exposed to the beam. Of these, there are three titanium capsules, which were sealed under helium atmosphere. An upstream capsule, 'US-Ti' was placed at the

5th location for all the irradiation phases. Two downstream capsules, 'DS-Ti1' and 'DS-Ti2', were placed at the most downstream 6th location during phase-1 and phase-3, respectively.

In the present work, four grades of dual α+β phase titanium alloy specimens in the DS-Ti1 capsule are provided for tensile and microstructural studies. The product summary and chemical composition for each grade are given in Table 1 and Table 2, respectively. Two grades of Ti-6Al-4V, standard Grade-5 (hereafter denoted as 'Gr-5') and Grade-23 Extra Low Interstitial, ELI ('Gr-23'), were procured in the shape of 1.6 mm-thick sheets in mill-annealed condition. The Gr-23 contains less oxygen (0.13% maximum) than the standard Gr-5, and improved ductility and fracture toughness are expected with some reduction in strength. A near-α phase α+β alloy Ti-3Al-2.5V Grade-9 ('Gr-9') was supplied in the shape of 0.7 mm-thick sheets. It contains less β-stabilizing vanadium element, exhibiting more α phase. It has less strength but more ductility than Ti-6Al-4V Gr-5/Gr-23. These sheet materials of Gr-5, Gr-23 and Gr-9 are annealed for 8 hours at 620 °C in a vacuum furnace. Another Grade-23 ELI, originally supplied as 12.5mm-ϕ round bar, was forged to around a 2 mm-thick sheet, and then annealed at 740 °C for 6 hours in a vacuum furnace ('Gr-23F'). By comparison with Gr-23 (sheet material), this may indicate different irradiation responses, if there is dependence on the thermomechanical treatments.

In the DS-Ti1 capsule, a variety of titanium alloy specimens were arranged into layers, as shown in Fig. 2. The upstream layer with 0.5mm thickness contains six tensile specimens, where the five gaps between them are filled with oval-shaped specimens for microstructural characterization. The tensile plate specimen has 42 mm-long dog bone shape with shoulder loading, with 0.5 mm-thick and 1.75 mm-wide gauge section. The downstream layer consists of 10 bend fatigue specimens with 1 mm thickness, which are to be tested at Fermilab by a newly-developed compact fatigue testing machine with remote handling capability [21]. Between these two layers, a 0.26 mm-thick Ti-6Al-4V Grade-23 ELI foil, with numerous small laser-machined cantilevers was installed for ultrasonic fatigue testing, an innovative technique invented at Oxford University [22]. The sample foil was manufactured from the center of the same 8-inch diameter bar, that was used to manufacture a J-PARC beam window. These macro-scale and meso-scale fatigue tests are expected to provide the first HCF data for irradiated Ti-6Al-4V alloy.

In the beam exposure at the BLIP facility, a raster beam with about 2.4 cm diameter footprint was delivered to the RaDIATE target box, with an average current of 154.0 µA and peak flux of $1.11 \times 10^{14}$ protons/(cm$^2$·s). The irradiation parameters for each phase are summarized in Table 3. The integrated numbers of delivered protons on the target box through the three irradiation phases reached $4.57 \times 10^{21}$ pot, which is more than typical pot for high intensity proton accelerator facilities per year. Table 4 is a summary of the irradiation on the three titanium capsules. After the phase-1 irradiation in 2017, the DS-Ti1 capsule experienced a total of $7.30 \times 10^{20}$ pot with peak fluence of $8.41 \times 10^{19}$ protons/cm$^2$. In Fig. 3(a) accumulated dpa on the DS-Ti1 specimens is shown as a function of radius, where dpa values were estimated with three models, the standard model by Norgett, Robinson, and Torrens (NRT) [23], NRT corrected with an efficiency function by Stoller et al [24], and NRT corrected with an efficiency function by Nordlund et al [25,26]. These efficiency functions take athermal recombination correction and atomic mixing caused by energetic cascade into account[2]. With the conventional NRT model, the peak damage is estimated to be 0.25 dpa, which is comparable to the existing typical irradiation data, as described in Sec. 1. Fig. 3(b) shows expected irradiation temperature distribution at the upstream tensile and microstructural specimen layer, estimated by Finite Element Analysis (FEA) simulation. There is a positive correlation between the accumulated dpa and the temperature, since the heat source is the energy deposition caused by proton beam flux. For 181 MeV proton at BLIP, the hydrogen (helium) gas production rate estimated by MARS code [27] are 120 (40) appm/dpa (NRT), which are fairly larger than those for the typical neutron irradiation, but are significantly smaller than those for the 30-120 GeV proton at J-PARC and Fermilab, about 300 (200) appm/dpa [28]. In Table 5, alloy grade, accumulated dpa (NRT), and expected irradiation temperature are summarized for each irradiated tensile and microstructural specimen.

---

[2] Recent measurements on displacement cross section by high energy proton beam exhibited remarkably good agreement to this improved model by Nordlund et al [52,53].

In the end of CY2017, after a few months of cool down, the DS-Ti1 capsule was sent to PNNL, together with another irradiated capsule containing silicon etc., in a US Department of Transportation (DOT) Type A container. The capsule opening and specimen sorting were conducted within a modular hot cell at the Radiochemical Processing Laboratory (RPL) at PNNL using a master-slave manipulator. Tensile tests were performed on the Instron 8801 mechanical load frame, being set up in another hot cell at RPL. To locate each tensile specimen in the load frame efficiently by the manipulator, a self-aligning fixture was designed and fabricated. The load frame is equipped with a Centorr vacuum furnace, capable of operating up to 1,100 °C. The tensile test matrix includes non-irradiated archive specimens for comparison to the irradiated DS-Ti1 capsule specimens. Both room temperature (RT) tests and elevated temperature tests at 200 ˚C, which is the estimated maximum operational temperature of the J-PARC primary beam window [3], were conducted. For all tensile testing, the crosshead speed was set to 2.5 μm/s, and change of the gauge section length was assumed to be identical to the displacement of the fixture crossheads.

Transmission Electron Microscopy (TEM) was employed to examine the microstructures for current work. To prepare TEM specimens, Tenupol jet electro-polishing was used on 3 mm diameter discs, punched from the microstructural specimens. Refinements of the polishing conditions were required to avoid the β-phase being preferentially polished. The sample discs were then studied with an atomic resolution analytical microscope JEOL ARM200CF at 200 keV electron beam energy, to find signatures of the radiation damage in both α and β phases, in comparison with the un-irradiated control specimens.

3. Results

*3.1 Tensile tests*

The stress-strain curves for Ti-6Al-4V Gr-5 specimens are plotted together in Fig. 4(a). Unirradiated control specimens exhibited good repeatability between multiple measurements at each test temperature ($T_{test}$=RT and 200 °C). For tests at elevated temperature ($T_{test}$=200 °C), a large loss of strength was identified with a slight increase in elongation, in comparison with those at RT, as expected in the material database [8]. At both test temperatures, modest work hardening occurs, and the fracture mode is ductile. On the other hand, the test at RT for the irradiated specimen F4C, located at the edge of the capsule (as shown in Fig. 3(b)) and thus minimally irradiated (0.06 dpa at 47 °C), indicated a remarkably large increase in the yield strength by more than 150 MPa. The yield strength was nearly identical to the ultimate tensile strength, with a prominent reduction in the uniform elongation compared to the control (from 6.68% to 0.78%). This material became unable to work harden to any significant extent, even with this low level of irradiation. Total elongation after the irradiation became about half of the control (from 16.0% to 9.06%), and seemingly suffered from plastic instability.

Fig. 4(b) shows stress-strain curves for Ti-6Al-4V Gr-23 ELI. The repeatability of the duplicated measurements for unirradiated control specimens is satisfactory at both test temperatures, though less so than for Gr-5. The higher temperature tests give lower total elongation than that at RT, and total elongations at the two test temperatures are both lower compared to each of the Gr-5 results. These observations on the control specimens seem to be in disagreement with the material database, and to be examined further by microstructural studies[3]. There are three irradiated specimens with two different dose levels, one for 0.06 dpa at 47 °C (T4C, same dose and irradiation temperature as the Gr-5 specimen F4C) and two for 0.23 dpa at 125 °C (T5 and T8). The tensile behavior at the RT on Gr-23 T4C exhibited even more drastic changes than that for Gr-5 F4C, *i.e.* a similar increase of the yield strength, but a smaller uniform elongation (0.18%) and no work hardening capability, a steeper drop of the stress in the plastic region, and a greater reduction of total elongation to less than half (from 13.8% to 6.37%). Two tests at RT for the different dose levels (T4C 0.06 dpa and T5 0.23 dpa) exhibited slight increase of the ultimate tensile strength (+33 MPa) and loss of total elongation (from 6.37% to 5.03%) for higher dose. Changes of these properties between 0.06 dpa and 0.23 dpa are thus not so significant as those observed between the control and 0.06 dpa. The total elongation at 200 °C (T8) was almost the same as that at RT (T5), exhibiting the same tendency as those for the control measurements.

The stress-strain curves for the near-$\alpha$ $\alpha+\beta$ alloy Ti-3Al-2.5V Gr-9 specimens tested at RT are plotted in Fig. 5(a). For the control measurements, the yield and tensile strength are lower than Ti-6Al-4V, while the uniform and the total elongations are much larger. In the specimen N5 irradiated to 0.22 dpa at 112 °C (similar dpa as Gr-23 high dose specimens T5 and T8), the increase in yield strength of about +290 MPa is more than a factor of 2 larger than that for Gr-23 (+134MPa). Although large decreases of both uniform and total elongations by irradiation were observed, modest work hardening capability still remains, with a larger ultimate tensile strength than the upper yield point and a uniform elongation of 3.21%. This is a clear difference to the tensile behavior observed for Ti-6Al-4V Gr-5 and Gr-23.

Fig. 5(b) shows stress-strain curves for Gr-23F, Grade-23 ELI forged from round bar, tested at RT. The measurements of the control samples showed scatter, not only in strength and elongation, but also in elastic modulus. They are most probably due to a difference in local material properties caused by the specific thermomechanical processing. Tensile behavior for the specimen F52 irradiated to 0.22 dpa at 112 °C looks similar to that for Gr-23 shown in Fig. 4(b), except for the appearance of an upper yield point, which was not observed for the sheet material. It shows no work hardening capability as for the other Ti-6Al-4V grades, suggesting that forging may not be beneficial for resistance to radiation damage.

The mechanical properties obtained by these tensile measurements are summarized in Tables 6 and 7, for the control archives and for the irradiated specimens, respectively. Note that the yield strength for the Gr-5, Gr-23 and irradiated Gr-23F samples is the 0.2% offset yield strength, while the yield strength for the Gr-9 and unirradiated Gr-23F samples is the upper yield strength. As shown in Fig. 5(a)(b), the latter samples exhibited a load drop upon yielding with a distinct upper and lower yield strength. The upper yield strength in these cases represents the onset of plastic deformation, and is more comparable to the 0.2% offset yield strength than the lower yield strength, which represents the onset of strain hardening after the load drop. A synopsis of the measured effects of irradiation on the properties for each material grade is summarized in Table 8. All grades of Ti-6Al-4V Gr-5, Gr-23, and Gr-23F lose ductility with about more than 90% reduction on uniform elongation: 88% for Gr-5 and even nearly 100% for Gr-23 and for Gr-23F. For Gr-23, the large reduction rate on the uniform elongation is regardless to the testing temperature. On the other hand, 71% reduction on uniform elongation for Gr-9 is less than those for Ti-6Al-4V grades. The hardening effect, *i.e.* changes on the yield strength and ultimate tensile strength, are more apparent for Gr-9 than Gr-5, Gr-23 or Gr-23F. It is to be emphasized that for the near-α α+β phase Ti-3Al-2.5V Gr-9 alloy, modest work hardening capability still remains after irradiation, while α+β Ti-6Al-4V grades have essentially none.

---

[3] Preliminary electron backscatter diffraction (EBSD) analysis exhibited a more elongated grain structure and misorientation boundaries which are not present in Gr-5, indicating residual cold or warm worked structure.

*3.2 Microstructure*

To investigate cause of the prominent loss of ductility specifically observed for the irradiated Ti-6Al-4V, microstructural investigations were performed on Gr-5 and Gr-23. As given in Table 5, accumulated radiation damage was 0.12 dpa for Gr-5 (specimen F5), and 0.22 dpa for Gr-23 (T4), respectively. TEM studies were conducted both on the α-phase and β-phase matrices, in comparison with those of unirradiated control specimens.

An optical micrograph on cross-section of the procured Gr-5 sheet material (after heat treatment as described in Table 1) is shown in Fig. 6(a), where a microstructure consisting of fine, equiaxed and acicular α-phase grains is observed. Figs. 6(b)(c) are micrographs of Gr-5 control specimen by scanning electron microscope (SEM) with backscattered electrons (BSE). It is a duplex microstructure with α phase grains and transformed β phase grains, with a few microns in size. For the dual-phase Ti-6Al-4V alloy, a wide variety of microstructures can be realized by applying optimized working and heat treatment processes [29]. The observed microstructure is typical for mill-annealed heat treatment condition followed by a hot-roll pressing [1]. On cooling to room temperature, the transformed β grains form α lamella by rejecting vanadium, a β stabilizer, giving some retained β phase in between the α variants. The microstructure for the Gr-23 specimen appeared to be quite similar to that of Gr-5.

Fig. 7(a) is a bright field (BF) image of an α-phase matrix for Gr-23 unirradiated control specimen with low magnification, where some grain boundaries are visible. Inside of each grain, a microstructure with a moderate dislocation density is identified, possibly resulting from either cold or hot working processes during fabrication. Fig. 7(b) is a BF image of an α-phase grain for the Gr-23 specimen T4 irradiated to 0.22 dpa at 120 °C. Many small spots with black dot contrast are identified everywhere, with less than 2 nm in size and number density of about $5 \times 10^{22}/m^3$, which do not appear in the unirradiated control specimen. Fig. 7(c) is a g/5g weak beam dark field (WBDF) image corresponding to the same region as given in Fig. 7(b), where numerous small white dot defects can be identified. These black/white dots, commonly formed in α phase grains, are most likely defect clusters, as typically identified in materials experiencing irradiation at low temperature. They are considered to be the main reason to cause the irradiation hardening observed in the tensile tests.

Fig. 8(a) is a BF image of Gr-5 unirradiated control specimen with low magnification, which shows a colony of mixed α and β matrices separated by grain boundaries. The size of β-phase grains is relatively smaller than that of main α-matrices. Fig. 8(b) is a close up on a β phase grain of the irradiated Gr-23 specimen T4. The black spots of defect clusters, which appeared in the α phase grains, are not identified in the β-phase grain at all, while a weak mottled contrast exists as background. These signatures are similar to a recent observation on microstructures for an irradiated metastable β phase titanium alloy with fine precipitation [7]. Fig. 8(c) is a diffraction pattern (DP) for this β-matrix in a $[110]_\beta$ zone axis (ZA), exhibiting pronounced extra spots at 1/3 and 2/3 $<112>_\beta$ positions, together with the parent β matrix reflections. The observed DP is typical for the ω phase precipitation in the metastable-β phase titanium and zirconium alloy systems, which can be found elsewhere in past publications [30–35]. These extra spots are consistent with the presence of the different ω–phase variants, which are formed by the displacements of atoms to <111> directions of mother β matrix, as expected for the displacive transformation from β (BCC) to ω (HCP) phase [36,37]. They can also be identified as distinct spots in the DP with $[113]_β$ ZA.

Figs. 9(a)-(c) are centered dark field images of the β phase grain of Gr-5/Gr-23 specimens, which are taken by using one of extra spots corresponding to the ω–phase reflection in each DP, as highlighted in subset. Fig. 9(a) is for the Gr-5 unirradiated control specimen, where DP exhibits diffuse reflection lines or diffuse streaks between the diffraction spots of the mother β matrix. Distinct extra spots of the ω phase are not identified in unirradiated condition (it is confirmed also for the Gr-23 control specimen). By choosing a brighter area in a diffuse streak, uniform distribution of very fine, nano-scale particles has been observed, with average size of 1.4 nm and number density of $1.5 \times 10^{24}/m^3$. These objects can be interpreted as precursors of the ω phase, as argued in a previous study [31]. A precursor/embryo is a perturbation of the BCC stability, and small-scale compositional fluctuation, displacement, and/or stress variation can favor the ω phase transformation, where the variations can either be random or due to β-phase split etc. The subset of Fig. 9(b) shows DP for the Gr-5 specimen F5, irradiated to 0.12 dpa at 72 °C. Compared to the unirradiated control (Fig. 9(a)), it apparently shows discrete extra spots and less diffuse reflection lines or diffuse streaks. The particles appeared in the centered dark field image exhibit very similar number density as for the unirradiated condition, while size of the particles seems to become coarser. It would mean the phase transformation from the precursors to the ω phase precipitates is induced by irradiation. Fig. 9(c) is for the Gr-23 specimen T4 irradiated up to 0.22 dpa at 120 °C, where diffuse streaks are very weak, and pronounced spots corresponding to the ω–phase appear instead. The number density of fine particles is similar or less to the other two conditions of Gr-5, which may be due to the difference of heat process between Gr-5 and Gr-23 materials. Meanwhile, size of the particles, about 2 nm in average, is apparently coarser than those observed in the Gr-5 specimens with none or less irradiation doses. Growth of the ω–phase formation is most likely in progress with increasing irradiation dose.

In Table 9, the number density and the average size of the particles in the β matrices, corresponding to the ω phase and/or its precursor, are summarized for each irradiation dose, together with the characteristics of defect clusters observed in the α matrices. Note that these are conservative estimates due to uncertainties in the measurements, since only part of multiple ω phase variants and subvariants is counted by using one of the extra spots, part of diffuse reflection lines or diffuse streaks, and some defect clusters in the α–grains are likely to be invisible. Even with the conservative estimation, the ω particle sizes appear to be very narrow and small, and the number densities are very high. The average sizes of the ω phase clearly indicate a tendency to increase during irradiation, or at least to be restructured from the fine particles (precursors) appeared in the unirradiated condition. It is known as the "ω-embrittlement" appeared in the metastable β titanium and zirconium alloys by ageing heat treatment, that causes increase in strength and especially large reduction in ductility due to coarsening of the fine precipitation of ω phase [38,39]. It is worth examining a possibility, that the radiation-induced ω-phase transformation in the β-matrices to be the cause of the prominent loss of ductility, observed specifically for dual phase Ti-6Al-4V alloy Gr-5/Gr-23 (as shown in Figs. 4(a)(b) and Fig. 5(b)), not for the near α Ti-3Al-2.5V Gr-9 with less β phase (Fig. 5(a)).

On the other hand, the size of the defect clusters (black dots) formed in the α-matrices by the irradiation is in the same nano-meter range but about 30 times lower in number density than the ω phase particles in the β-matrices. These sizes and number densities of defect clusters, however, can still lead to significant irradiation hardening for the α-matrices, which might be a reason to cause the larger increase of the yield strength observed for Ti-3Al-2.5V Gr-9.

## 4. Discussion

In the TEM analyses on dual phase Ti-6Al-4V alloys, we have identified direct evidence of the ω phase transformation in the β matrices, and indicated its possible coarsening and/or restructuring during proton beam irradiation[4]. So far, the ω phase was studied intensively in metastable β (BCC)-phase titanium and zirconium alloy systems, and is known to be formed both during quenching (rapid cooling from high temperature) and during ageing heat treatment below α-phase forming temperature. It is generally agreed that the isothermal ω phase, which forms during ageing, can be described by a conventional nucleation and growth process[5]. A phase separation could occur at the ageing temperature accompanied by rejection of solute from ω to β, and the ω phase could form in the solute poor region [38,40]. On the other hand, the mechanism to form the athermal ω phase during quenching is believed to be a diffusionless displacive transformation. This phase forms through the consecutive collapse of pairs of $\{111\}_\beta$ planes, as a result of a frozen soft phonon mode in the β phase [36,41]. The formed particles, either the trigonal ω phase or the hexagonal ω phase, depending on the extent of planar collapse, are compositionally indistinct from the surrounding matrix. On isothermal annealing at low temperatures, these athermal ω precipitates coarsen to form isothermal ω precipitates with more ordered compositional changes and morphology. As a result, the isothermal ω appears as sharper spots in the diffraction pattern, whereas the athermal ω phase appears as diffuse streaks. This could be due to significant amounts of perturbation on the atom columns, grouping of collapse, and/or directionality of the one-dimensional defects which have 12 variants. It could also be linked to local compositional fluctuations, possibly being treatable with a pseudo-spinodal concept [42]. The gradual changes in diffraction pattern from diffuse streaks to distinct spots of the ω-phase reflections, shown in Fig. 9(a)-(c), look quite similar to what was demonstrated for transition process from the athermal ω to isothermal ω under ageing [43]. High-intensity proton irradiation produces a high density of vacancies and interstitials, and clusters of these defects, potentially leading to partition the atoms to the precursors, which then continue to initiate the ω phase transformation. The enhanced diffusion from these defects, albeit at low temperature, might allow the ω phase to coarsen, similar to the isothermal ω phase formation under ageing. The vacancies and interstitials could also affect to the soft phonon mode in the β phase and promote the planner collapse of the $\{111\}_\beta$ to coarsen the ω phase. In fact, there are arguments that point defects and interstitials may play an essential role in stabilizing ω [36,44], and there is a former observation reporting irradiation with 1 MeV electrons causing the ω phase in a Zr-Nb alloy to coarsen with increasing irradiation dose, which is attributed to the enhanced diffusion [45]. To further investigate the isothermal-like ω phase formation, it is worth performing elemental analyses, such as energy dispersive X-ray spectroscopy (EDX), electron energy-loss spectroscopy (EELS), and atom probe tomography (APT).

In former works by one of authors, a Ti-35Al-10V intermetallic compound, manufactured by powder metallurgy, was examined with neutron irradiation [46,47]. In unirradiated condition, it exhibits remarkably high ductility, more than 60% elongation at 600 °C. The alloy contains β phase with ordered BCC structure, considered to play an important role for this plastic behavior. However, after the neutron irradiation of $3.5 \times 10^{25}$ n/cm$^2$, the total elongation decreases to only 10%. The result of the tensile test suggested that the embrittlement was caused by phase decomposition during irradiation, and formation of ω phase in β phase was anticipated as its cause [46][6].

---

[4] The TEM identification of ω phase was firstly made in February 2019 at PNNL, and efforts to confirm its existence was launched since then. An independent study by Energy dispersive X-ray diffraction (EDXRD) is in progress by one of the authors at BNL NSLS II [54].

[5] There is a recent work reporting diffusionless isothermal ω transformation in Ti-V alloys driven by quenched-in compositional fluctuations [55].

For the irradiated Gr-5/Gr-23 microstructural specimen, there is a high density of small, nanoscale defect clusters forming in the α phase, that likely will lead to noticeable hardening and reduction of the ductility in itself. There is no easy way to separate out the effect of ω-hardened/embrittled β phase versus defect-hardened α phase from the current data. As shown in Fig. 6, the β phase is a small volume fraction compared to the α phase. A preliminary EBSD analysis on the Gr-23 specimen gives the phase composition to be about 96% for α and 4% for β. Although the fraction of the β phase matrices appeared to be relatively small, they disperse at grain boundaries of the α matrices, and the overall mechanical properties, especially final fracture behavior, might be potentially affected by the presence of the hardened/embrittled β-matrices due to the ω phase formation and coarsening. It is worth examining further the fracture surfaces and necked regions and/or TEM investigations on the deformed specimens. Atomic force microscope (AFM) in conjunction with the EBSD is now in progress to evaluate hardness for each of α and β phase grain before/after irradiation.

In a former work by part of authors [7], we discussed possible radiation damage tolerance of a metastable β phase alloy Ti-15V-3Cr-3Sn-3Al (15-3 Ti) owing to its fine precipitation. We studied thin solution-treated 15-3 Ti foil without ageing, irradiated up to 0.12 dpa by 30 GeV high-energy proton at J-PARC. In the TEM analysis, we observed thick diffuse streaks in DP (similar pattern as Fig. 9(a), but much brighter diffuse streaks), and rich nano-scale precipitation in the centered dark field image with density greater than $10^{23}/m^3$. They were identified as either orthorhombic martensite α" phase or athermal ω phase precipitation. It is to be noted, that these diffuse streaks in DP stayed the same between unirradiated and irradiated conditions, and micro-Vickers indentation tests suggested almost no hardening occurred by the irradiation. It is of great contrast to the present work, where the β-matrices of Ti-6Al-4V exhibited more discrete extra spots of the ω-phase, which were coarsened with higher irradiation dose, and tensile data showed large radiation hardening only with 0.06 dpa. With respect to the development of next-generation radiation-tolerant beam window material, this contrasting behavior against the irradiation between the dual α+β phase Ti-6Al-4V (current beam window material) and the metastable-β phase 15-3 Ti is quite worth studying further. For that purpose, we are conducting ion beam irradiation experiments on these two alloys, to be examined by nano-indentation hardness testing and microstructural investigations [48]. It is also to be noted that tensile, microstructural, and meso-fatigue foil specimens of the 15-3 Ti alloy were installed in the DS-Ti2 capsule for current BLIP irradiation campaign, and already irradiated up to 0.95 dpa as shown in Table 4. Tensile and microstructural studies are now in progress on these specimens, and to be reported as separate works.

## 5. Conclusion

A high-intensity proton beam exposure with 181 MeV energy has been conducted at BNL BLIP facility on various material specimens for accelerator targetry applications. They include titanium alloys as a beam window material, separating accelerator vacuum and target station vessel atmosphere. Out of three titanium capsules, the radiation damage level of the analyzed

---

[6] However in the later publication [47], TEM studies could not confirm the radiation induced ω phase formation. A DP of ω phase formation in $[113]_β$ ZA, similar to Fig. 9(c) was obtained for (unirradiated) specimen, aged at very high temperature.

DS-Ti1 capsule was 0.25 dpa (NRT) at beam center region with an irradiation temperature around 120 °C. Tensile tests showed increased hardness and a large decrease in ductility for the dual α+β-phase Ti-6Al-4V Gr-5 and Gr-23 ELI alloys, with the near α-phase Ti-3Al-2.5V Gr-9 alloy still exhibiting uniform elongation of a few % after irradiation. TEM analyses on Ti-6Al-4V indicated clear evidence of a high-density of defect clusters with size less than 2 nm in each α-phase grain. The β-phase grains did not contain any visible defects such as loops or black dots, while the diffraction patterns clearly indicated the ω-phase in an advanced formation stage. It indicates that the radiation-induced ω-phase transformation in the β-phase could lead to greater loss of ductility in Ti-6Al-4V alloys in comparison with Ti-3Al-2.5V alloy with less β-phase. Further investigations, including EBSD, AFM and XRD, will continue to evaluate these arguments in more quantative manner.

While the industry standard Ti-6Al-4V alloy appears favorable for use in particle accelerator beam windows due to its high strength and reasonable ductility, it appears highly susceptible to irradiation damage. The main issue of concern is a severe loss of ductility and work hardening capability at radiation doses as low as 0.06 dpa. This can be a result of different radiation damage effects on the α and β-phase grains. Near single phase Ti-3Al-2.5V, while not as strong initially as Ti-6Al-4V, retains more ductility and some work hardening capability at doses of up to 0.22 dpa. Systematic studies will continue on various titanium alloys in the other capsule, including metastable β-phase 15-3 Ti with rich nano-scale precipitates as defect sink sites [7] and other radiation-tolerant candidate grades such as ultra-fine-grained Ti-6Al-4V [49], to determine future beam window materials in US and Japan high-intensity proton accelerator facilities.

**Acknowledgement**

This work is financially supported by the U.S.-Japan Science and Technology Cooperation Program in High Energy Physics.

This manuscript has been authored by Fermi Research Alliance, LLC under Contract No. DE-AC02-07CH11359 with the U.S. Department of Energy, Office of Science, Office of High Energy Physics.

This document was prepared by RaDIATE collaboration using the resources of the Fermi National Accelerator Laboratory (Fermilab), a U.S. Department of Energy, Office of Science, HEP User Facility. Fermilab is managed by Fermi Research Alliance, LLC (FRA), acting under Contract No. DE-AC02-07CH11359.

The authors would like to note the contributions of our colleague, Senior Scientist Emeritus Nikolaos Simos at BNL, who passed away 15 July 2020 unexpectedly. Dr. Simos initiated and made possible the pioneering contributions to radiation damage studies of high power targetry materials, which became the foundational basis of activities conducted by the RaDIATE collaboration for over a decade.

**Table 1**
Product summary of the four tested titanium alloy grades in the DS-Ti1 capsule.

|  | Vendor | Industrial Standards | Shape | Heat Treatments |
|---|---|---|---|---|
| Gr-5 | TIMETAL 6-4 | AMS 4911 M<br>ASTM B 265 Grade-5 | 0.063in (1.6 mm)-thick sheet<br>mill-annealed | annealed for 8 h at 620°C in vacuum furnace |
| Gr-23 | TIMETAL 6-4 Extra Low Interstitial (ELI) | AMS 4907 H<br>ASTM B 265 Grade-23 | | |
| Gr-9 | NSSMC | (JIS H 4600 61)<br>(ASTM B 265 Grade-9) | 0.7 mm-thick sheet | |
| Gr-23F | DAIDO | (ASTM B 265 Grade-23) | 12.5mm-$\phi$ bar, forged to ~2mm-thick sheet | annealed for 6 h at 740°C in vacuum furnace |

**Table 2**
Chemical composition [wt.%] of each grade.

|  | C | O | N | H | Fe | Al | V | Others |
|---|---|---|---|---|---|---|---|---|
| Gr-5 | 0.017 | 0.19~0.20 | 0.08~0.09 | - | 0.16~0.15 | 6.35~6.27 | 3.99~4.00 | <0.0005 |
| Gr-23 | 0.008 | 0.13~0.11 | 0.004~0.002 | - | 0.12~0.15 | 6.15~6.09 | 3.85~4.00 | <0.0004 |
| Gr-9 | 0.01 | 0.076 | 0.007 | 0.0028 | 0.05 | 2.90 | 2.41 | - |
| Gr-23F | 0.009 | 0.10 | 0.005 | 0.0046 | 0.22 | 5.99 | 4.23 | - |

**Table 3**
Summary of the three irradiation phases during 2017–2018 on the RaDIATE target box at the BLIP facility.

|  |  | 2017 | | 2018 | Total |
|---|---|---|---|---|---|
|  |  | Phase-1 | Phase-2 | Phase-3 |  |
| Total Hours |  | 226.3 | 302.9 | 789.1 | 1,318.3 |
| Total Days |  | 9.4 | 12.6 | 32.9 | 54.9 |
| Average current | ($\mu$A) | 143.5 | 150.6 | 158.4 | 154.0 |
| Peak flux | (p/cm$^2$/s) | 1.03E+14 | 1.08E+14 | 1.14E+14 | 1.11E+14 |
| Peak fluence | (p/cm$^2$) | 8.41E+19 | 1.18E+20 | 3.23E+20 | 5.25E+20 |
| pot |  | 7.30E+20 | 1.03E+21 | 2.81E+21 | 4.57E+21 |

**Table 4**
Summary of irradiation on three titanium capsules.

|  | Position | Irradiation Phases | Peak fluence (p/cm$^2$) | Pot | Total peak dpa (NRT) |
|---|---|---|---|---|---|
| DS-Ti1[*] | 6 | 1 | 8.41E+19 | 7.30E+20 | 0.25 |
| DS-Ti2 | 6 | 3 | 3.23E+20 | 2.81E+21 | 0.95 |
| US-Ti | 5 | 1,2,3 | 5.25E+20 | 4.57E+21 | 1.53 |

[*]: In the current work, specimens in downstream (DS)-Ti1 capsule irradiated in the phase-1 have been analyzed.

**Table 5**
Specimen ID, alloy material grade, accumulated dpa (NRT), and expected irradiation temperature ($T_{irr}$) for tensile and microstructural specimens in the DS-Ti1 capsule.

| tensile | material | D (dpa) | $T_{irr}$ (°C) | micro | material | D (dpa) | $T_{irr}$ (°C) |
|---|---|---|---|---|---|---|---|
| T4C | Gr-23 | 0.06 | 47 | F5 | Gr-5 | 0.12 | 72 |
| F52 | Gr-23F | 0.22 | 112 | F6[*] | Gr-5 | 0.25 | 129 |
| T5 | Gr-23 | 0.23 | 125 | T4 | Gr-23 | 0.22 | 120 |
| T8 | Gr-23 | 0.23 | 125 | N2[*] | Gr-9 | 0.25 | 129 |
| N5 | Gr-9 | 0.22 | 112 | F7 | Gr-5 | 0.12 | 72 |
| F4C | Gr-5 | 0.06 | 47 |  |  |  |  |

[*]: These two microstructural specimens were provided for a thermal shock experiment at CERN HiRadMat facility [20].

**Table 6**
Summary of the tensile measurements on unirradiated control specimens, for yield strength (YS), ultimate tensile strength (UTS), uniform elongation (UE), and total elongation (TE).

| Unirr. | $T_{test}$=RT | | | | $T_{test}$=200°C | | | |
|---|---|---|---|---|---|---|---|---|
| | YS (MPa) | UTS (MPa) | UE (%) | TE (%) | YS (MPa) | UTS (MPa) | UE (%) | TE (%) |
| Gr-5 | 1,044 | 1,094 | 6.68 | 16.0 | 768 | 861 | 9.09 | 18.9 |
| Gr-23 | 1,012 | 1,091 | 8.98 | 13.8 | 792 | 891 | 6.31 | 11.2 |
| Gr-9 | 614 | 729 | 11.2 | 26.4 | - | - | - | - |
| Gr-23F | 902 | 915 | 5.50 | 11.8 | - | - | - | - |

**Table 7**
Summary of the tensile measurements on irradiated DS-Ti1 capsule specimens, for yield strength (YS), ultimate tensile strength (UTS), uniform elongation (UE), and total elongation (TE).

| Irrad. | dpa | $T_{irr}$ (°C) | $T_{test}$=RT | | | | | $T_{test}$=200°C | | | |
|---|---|---|---|---|---|---|---|---|---|---|---|
| | | | Spec. ID | YS (MPa) | UTS (MPa) | UE (%) | TE (%) | Spec. ID | YS (MPa) | UTS (MPa) | UE (%) | TE (%) |
| Gr-5 | 0.06 | 47 | F4C | 1,200 | 1,204 | 0.78 | 9.06 | - | - | - | - | - |
| Gr-23 | 0.06 | 47 | T4C | 1,146 | 1,147 | 0.18 | 6.37 | - | - | - | - | - |
| | 0.23 | 125 | T5 | 1,170 | 1,180 | 0.13 | 5.03 | T8 | 949 | 949 | 0.19 | 5.05 |
| Gr-9 | 0.22 | 112 | N5 | 901 | 946 | 3.21 | 10.9 | - | - | - | - | - |
| Gr-23F | 0.22 | 112 | F52 | 1,059 | 1,074 | 0.06 | 6.25 | - | - | - | - | - |

**Table 8**
Synopsis of tensile property changes caused by the irradiation, for yield strength (YS), ultimate tensile strength (UTS), uniform elongation (UE), and total elongation (TE).

| Irrad./Unirr. | dpa | $T_{irr}$ (°C) | $T_{test}$=RT | | | | $T_{test}$=200°C | | | |
|---|---|---|---|---|---|---|---|---|---|---|
| | | | YS (%) | UTS (%) | UE (%) | TE (%) | YS (%) | UTS (%) | UE (%) | TE (%) |
| Gr-5 | 0.06 | 47 | +15 | +10 | -88 | -43 | - | - | - | - |
| Gr-23 | 0.06 | 47 | +13 | +5 | -98 | -54 | - | - | - | - |
| | 0.23 | 125 | +16 | +8 | -99 | -64 | +20 | +7 | -97 | -55 |
| Gr-9 | 0.22 | 112 | +47 | +30 | -71 | -59 | - | - | - | - |
| Gr-23F | 0.22 | 112 | +17 | +17 | -99 | -47 | - | - | - | - |

**Table 9**
(a) The measurements on the ω phase particle/precursor size and density in the β matrices of unirradiated Gr-5, irradiated Gr-5 and Gr-23; (b) The defect cluster density in the α matrices for the irradiated Gr-5 and Gr-23.

| | dpa | (a) ω Particle/Precursor Characteristics | | (b) Defect Cluster Characteristics | |
|---|---|---|---|---|---|
| | | Avg. Size(nm) | Density(/m$^3$) | Avg. Size(nm) | Density(/m$^3$) |
| Gr-5 Unirr. | - | 1.4 | 1.5E+24 | - | - |
| Gr-5 Irrad. | 0.12 | 1.6 | 1.6E+24 | < 2 | 5.2E+22 |
| Gr-23 Irrad. | 0.22 | 2.1 | 8.0E+23 | < 2 | 3.8E+22 |

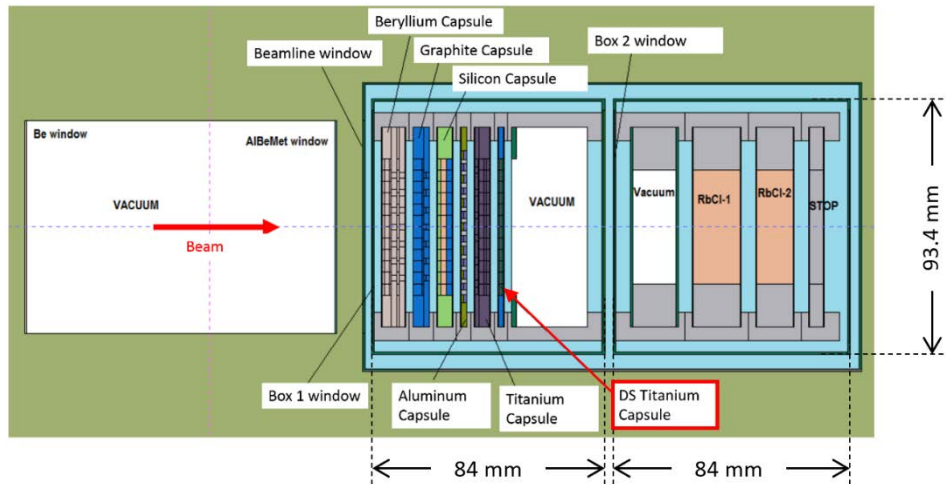

Fig. 1. Top view of the target box arrangement at the BLIP beam line. The RaDIATE target box (Box 1) was placed at the upstream of the isotope production target box (Box 2). The downstream titanium capsule (DS-Ti1) for current work was placed at the most downstream of the 6 capsules in Box 1, in front of a thick vacuum degrader capsule.

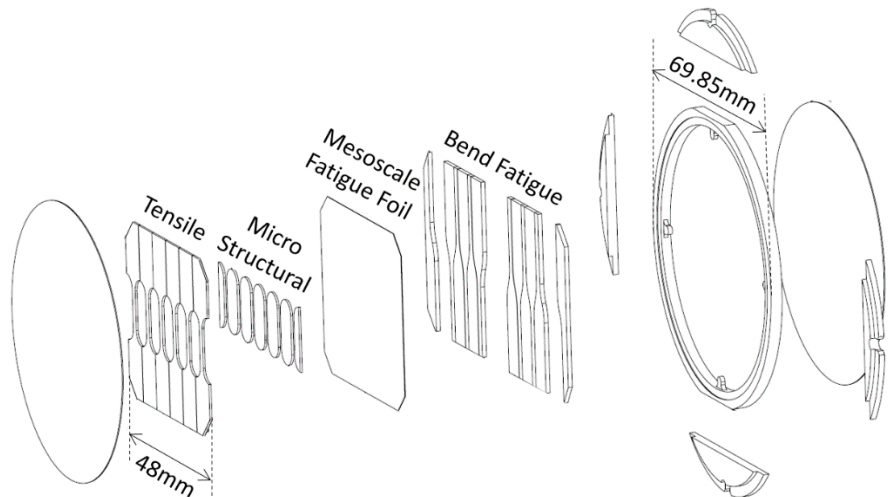

Fig. 2. Exploded view of the DS-Ti1 capsule assembly. The tensile and microstructural specimens for current work were placed at the most upstream.

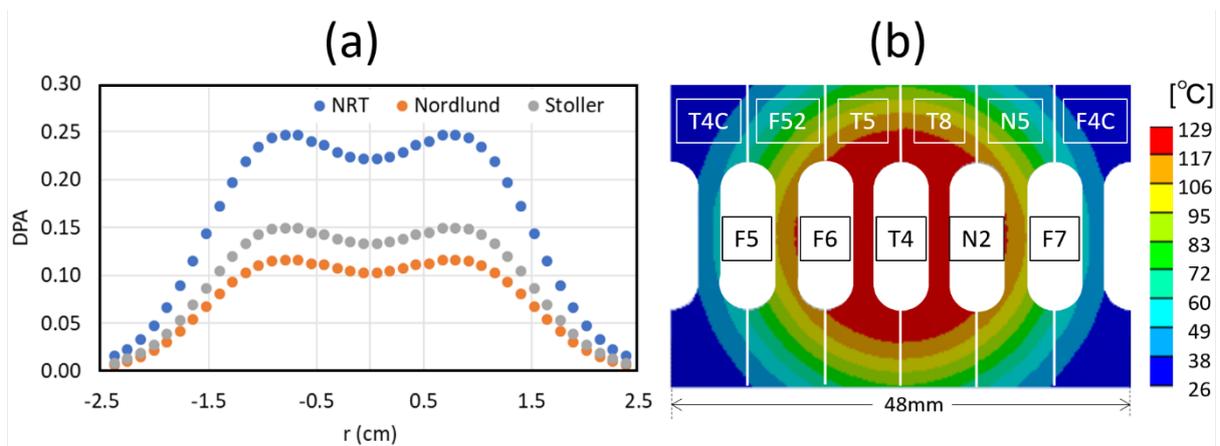

Fig. 3. (a) Accumulated dpa as function of radius, estimated by three different models; (b) Expected irradiation temperature distribution at the upstream tensile/microstructural specimen layer for the DS-Ti1 capsule, estimated by simulation.

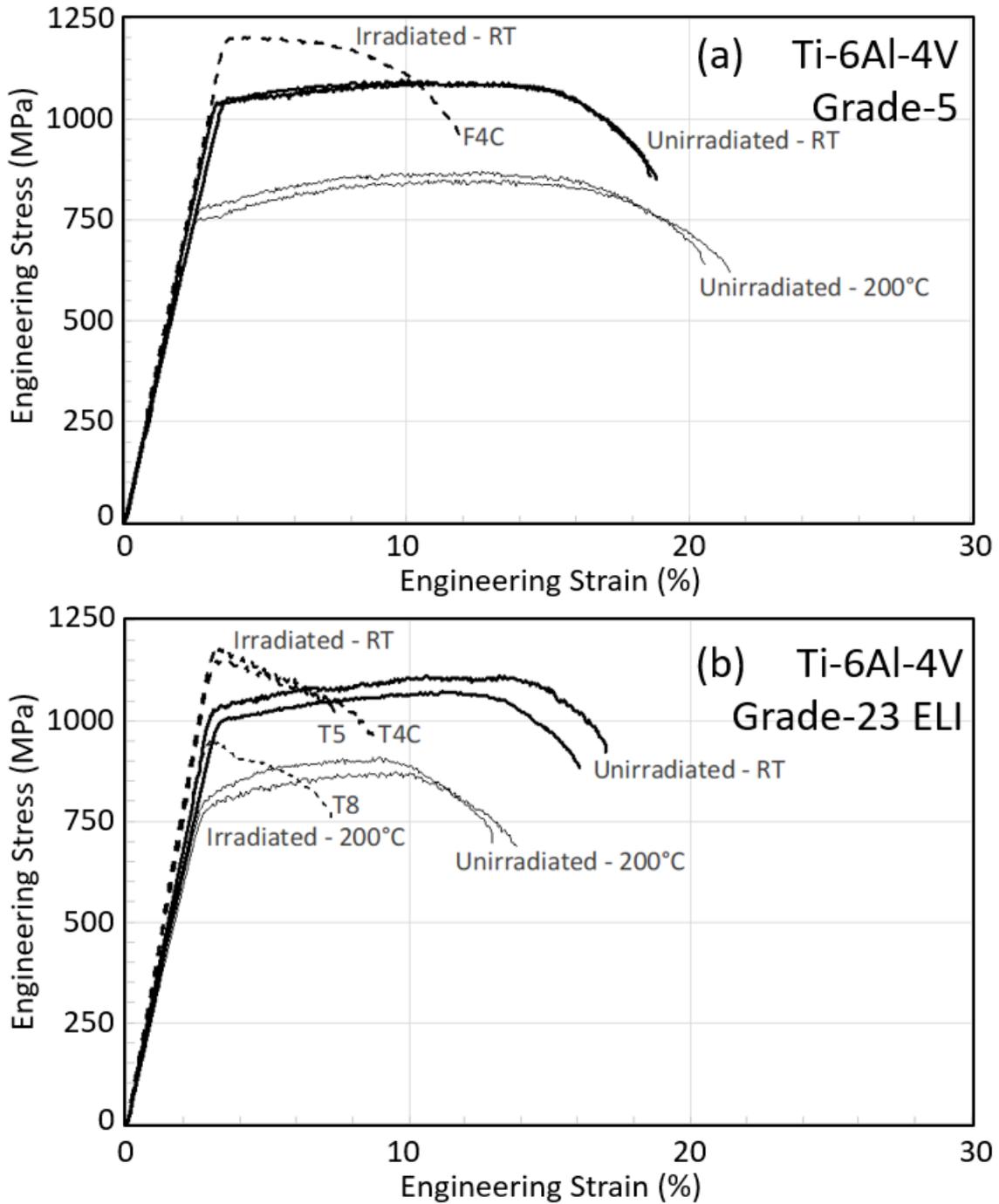

Fig. 4 Stress-strain curves for α+β alloy Ti-6Al-4V. Solid lines for unirradiated specimens, dashed lines for irradiated specimens, thick lines for tests at room temperature (RT), and thin lines for tests at 200 °C: (a) Grade-5, F4C=0.06 dpa at $T_{irr}$=47 °C; (b) Grade-23 ELI, T4C=0.06 dpa at $T_{irr}$=47 °C, T5/T8=0.23 dpa at $T_{irr}$=125 °C.

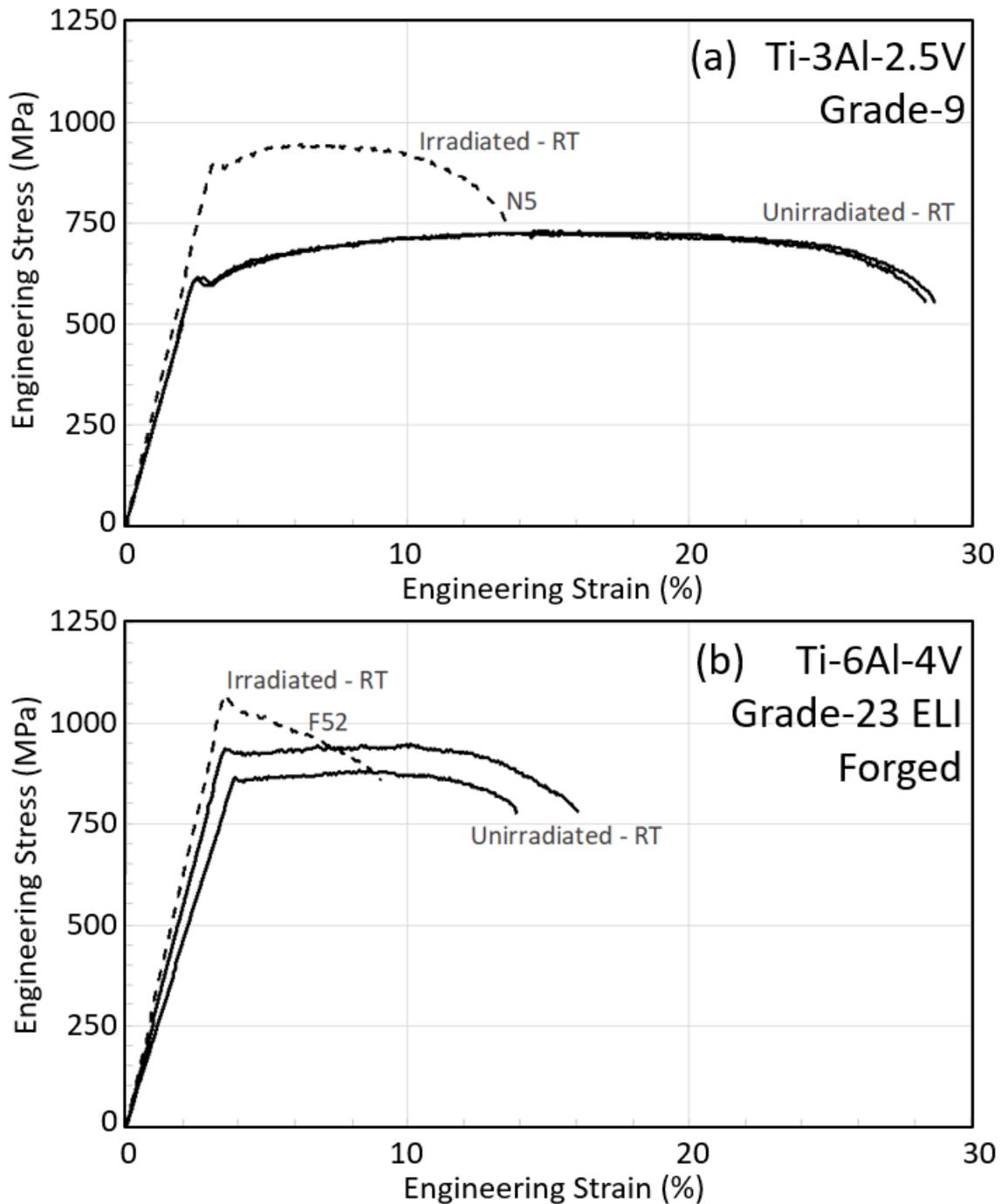

Fig. 5 Stress-strain curves at room temperature (RT). Solid lines for unirradiated specimens and dashed lines for irradiated specimens: (a) α-like α+β alloy Ti-3Al-2.5V Grade-9, N5=0.22 dpa at $T_{irr}$=112 °C; (b) α+β alloy Ti-6Al-4V Grade-23 ELI, fabricated by forging from round bar, F52=0.22dpa at $T_{irr}$=112 °C.

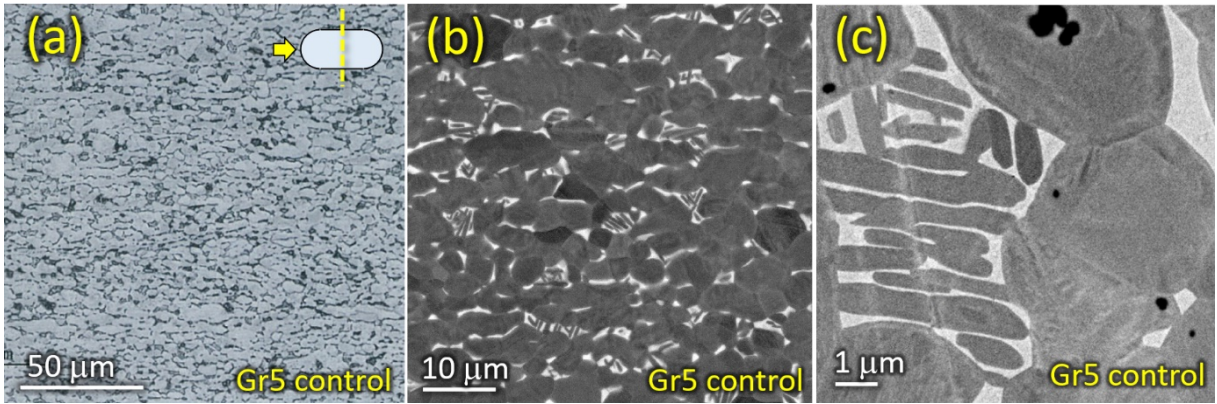

Fig. 6. Micrographs of Ti-6Al-4V Grade-5 for cross section of the sheet material: (a) An optical image and (b)(c) BSE-SEM images. A duplex microstructure of equiaxed α-phase grains surrounded by small amount of transformed inter-granular β phase regions is identified.

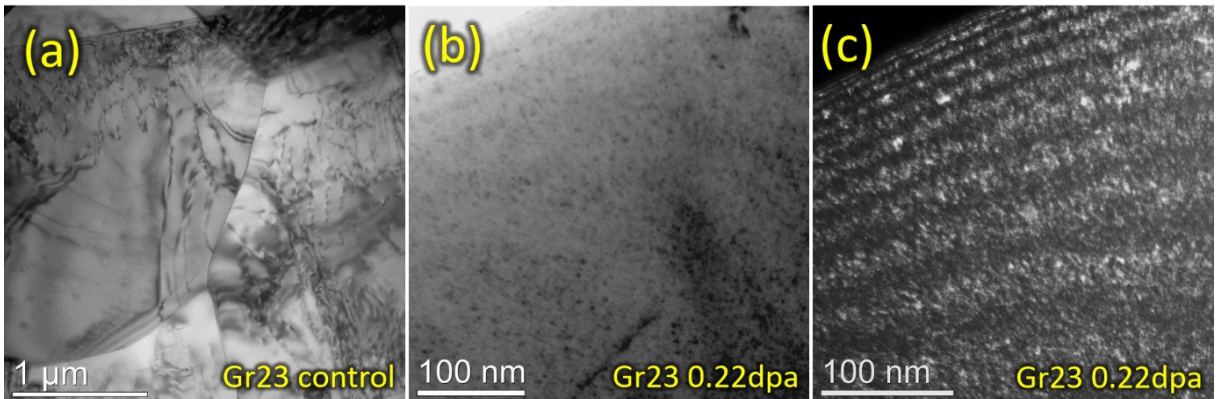

Fig. 7. TEM observations on the α-phase grains of Ti-6Al-4V Grade-23 ELI specimens: (a) BF image of unirradiated control. (b) BF image for the microstructural specimen T4 irradiated to 0.22 dpa. (c) g/5g WBDF image for the same region as (b). For (b)(c), high density small black/white dots, likely defect clusters of less than 2 nm in size, are identified.

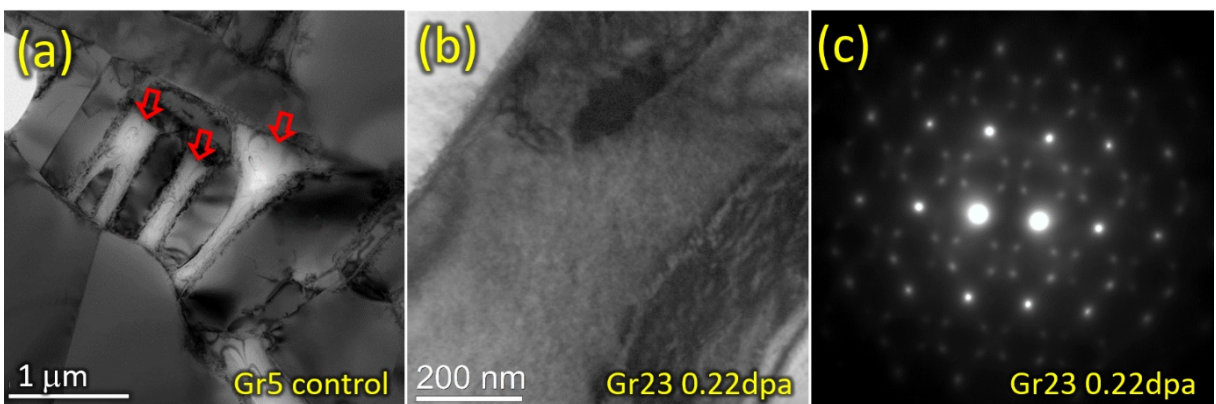

Fig. 8. TEM observations on the β-phase grains of Ti-6Al-4V: (a) BF image of Grade-5 unirradiated control specimen, exhibiting a colony of mixed α/β phases, with β phases indicated by arrows. (b) BF image for a β phase grain of Grade-23 ELI specimen T4 irradiated to 0.22 dpa. Mottled background suggests existence of fine precipitation. (c) DP for the same region as (b) with $[110]_\beta$ ZA. Pronounced extra spots corresponding to the ω–phase are observed.

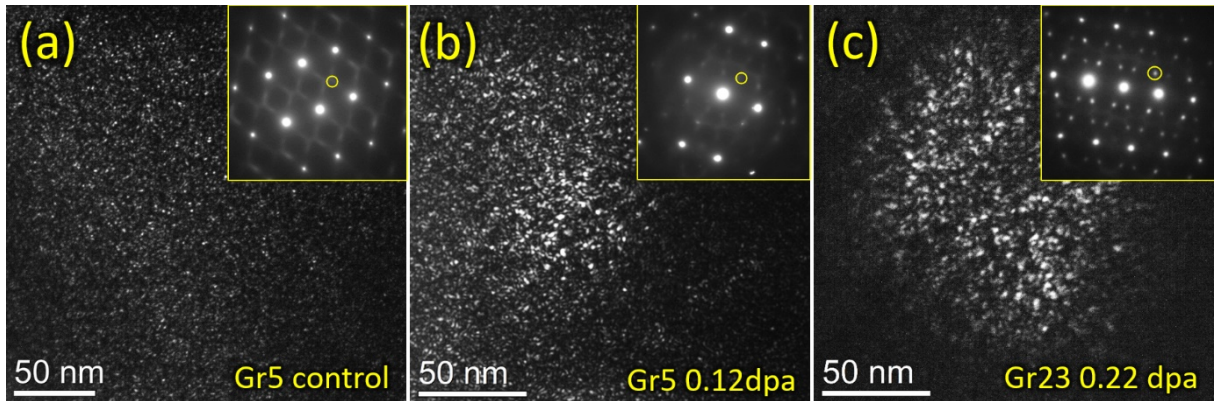

Fig. 9. Comparison of the ω–phase / precursor distribution in Ti-6Al-4V β-matrices for different irradiation levels, where centered dark field images are taken using a reflection highlighted with circle in DP as shown in each subset: (a) The Grade-5 unirradiated control. A DP with $[110]_\beta$ ZA shows diffuse reflection lines or diffuse streaks, possibly from precursors. (b) Grade-5 specimen F5 irradiated to 0.12 dpa. A DP with $[113]_\beta$ ZA shows discrete spots with less diffuse streaks, to be interpreted as start of the ω–phase formation. (c) Grade-23 ELI specimen T4 irradiated to 0.22 dpa. Pronounced spots for ω–phase observed in a DP with $[113]_\beta$ ZA, with apparently coarser particles in the centered dark field image.